**GlycoPy: A CasADi-based Python Framework for**

**Hierarchical Modeling, Optimization, and Control of Bioprocesses**

Yingjie Ma[1,2], Jing Guo[2,3], and Richard D. Braatz[2,*]

[1]Nanjing University, 215163, Suzhou, Jiangsu Province, China.

[2]Massachusetts Institute of Technology, 02139, Cambridge, MA, USA.

[3]Polytechnique Montréal, H3T 0A3, Montréal, QC, Canada.

**Abstract**: Efficient implementation of nonlinear model predictive control (NMPC) for bioprocesses remains challenging because large nonlinear models are difficult to organize, simulate, and embed within optimization and control workflows. This difficulty is particularly pronounced for large-scale and multiscale systems that require hierarchical model construction and customized simulation strategies. To address this issue, we present GlycoPy, a CasADi-based Python framework for hierarchical modeling, optimization, and control of bioprocesses. GlycoPy combines an equation-oriented, object-oriented modeling architecture with CasADi's symbolic and differentiable computational capabilities, enabling hierarchical model composition, numerical and symbolic simulation, parameter estimation, dynamic optimization, and NMPC within a unified workflow. A key feature of the framework is its support for customized differentiable simulation algorithms that can be embedded directly in gradient-based optimization and control. GlycoPy is demonstrated on a multiscale monoclonal antibody glycosylation process in Chinese hamster ovary cell culture, where it is used for hierarchical model construction, quasi-steady-state simulation, and adaptive NMPC. The results show that GlycoPy provides a practical and reusable framework for applying advanced optimization and control methods to computationally demanding bioprocesses.

---

[*] To whom correspondence should be addressed: braatz@mit.edu.

## 1. Introduction

Model predictive control (MPC) is widely used in process industries (Qin and Badgwell, 2003) due to its ability to handle input and state constraints, manage multivariate systems, and optimize overall performance criteria (Köhler et al., 2024). Most existing applications employ linear models derived from empirical model identification to avoid the complexities of nonlinear and nonconvex optimization problems (Hedengren et al., 2014). Many bioprocess models are inherently nonlinear, however, and the limited predictive accuracy of linear input–output models restricts their applicability for dynamic processes that span a wide range of operating conditions (Huang et al., 2009). This limitation has spurred extensive research into applying MPC based on nonlinear first-principles models, aka nonlinear model predictive control (NMPC) (Espinel-Ríos et al., 2022; Hermanto et al., 2009; Nagy et al., 2007). Advances in control and optimization algorithms, as well as computational power, have alleviated concerns over computational time for many applications (Diehl et al., 2002; Zavala and Biegler, 2009). However, practical deployment remains challenging because large nonlinear bioprocess models are expensive to develop, maintain, and integrate into reusable estimation-and-control workflows (Lucia et al., 2017).

Significant progress has been made in software for dynamic optimization and nonlinear model predictive control over the past two decades. Frameworks such as MUSCOD-II (Leineweber et al., 2003), ACADO (Houska et al., 2011), GRAMPC (Englert et al., 2019), and acados (Verschueren et al., 2022) enable the efficient formulation and solution of dynamic optimization problems, which are solved repeatedly in NMPC and moving horizon estimation (MHE). To further streamline controller development, do-mpc provides integrated modules for simulation, estimation, NMPC, and uncertainty handling (Fiedler et al., 2023). However, do-mpc does not support hierarchical modeling, which limits its convenience for developing large-scale bioprocess models. To address this issue, equation-oriented environments with hierarchical modeling capabilities, such as Pyomo (Hart et al., 2017), have also been extended toward NMPC applications. For example, (Lozano Santamaría and Gómez, 2016) developed an NMPC framework based on Pyomo. Although simultaneous methods have strong theoretical properties for dynamic optimization, feasible-path approaches can be particularly attractive for some strongly nonlinear problems, because linearizations at infeasible points may lead to poor Newton steps and inaccurate multiplier estimates (Biegler, 2010). For this reason, control vector parameterization (CVP), which is available in software such as gPROMS (Siemens Process Systems Engineering Ltd., 2025), remains a useful design choice for the class of applications considered in this work. However, these frameworks typically rely on off-the-shelf differential-algebraic equation (DAE) solvers within the CVP iterations and do not readily support the development of customized differentiable simulation algorithms that can accelerate simulation by orders of magnitude while remaining suitable for gradient-based optimization.

CasADi provides a strong foundation for such a framework (Andersson et al., 2019). Its symbolic architecture allows users to compose multiple operations, including numerical integration, equation solving, and derivative evaluation, into a single differentiable computational graph. In addition to symbolic manipulation and algorithmic differentiation (AD), CasADi also offers numerical integration and nonlinear optimization capabilities, which make it well suited for developing dynamic optimization and NMPC algorithms with substantial flexibility. However, CasADi mainly provides low-level building blocks rather than a process-oriented framework for hierarchical model development and reusable control workflows. In particular, when dealing with large-scale and multiscale bioprocesses, users still need substantial additional effort to organize variables and equations across submodels, manage simulation inputs and outputs, and develop customized differentiable simulation algorithms that can be embedded efficiently in optimization and control. Although many software packages based on CasADi have been developed to streamline optimal control and NMPC (Bhonsale et al., 2016; Fiedler et al., 2023; Verschueren et al., 2022), they do not focus on hierarchical modeling or customized differentiable simulation. These needs motivate the development of GlycoPy, which combines the flexibility of CasADi with a hierarchical modeling architecture and workflow-oriented tools for simulation, optimization, and NMPC.

Therefore, we developed GlycoPy, a CasADi-based Python framework for hierarchical modeling, optimization, and control of bioprocesses. GlycoPy combines an equation-oriented, object-oriented modeling architecture with the symbolic and differentiable computational capabilities of CasADi, enabling users to focus on model construction while retaining the flexibility needed for advanced algorithm development. In addition to supporting hierarchical model composition, GlycoPy provides reusable workflows for simulation, parameter estimation, dynamic optimization, and nonlinear model predictive control. A particular strength of the framework is that it supports symbolic simulation and the construction of customized differentiable simulation algorithms, which is important for embedding efficient simulation strategies within gradient-based optimization and control. In this work, dynamic optimization in GlycoPy is implemented using CVP, which aligns naturally with the simulation-oriented workflow and the class of strongly nonlinear bioprocess applications considered here.

The development of GlycoPy was motivated by our work on modeling, optimization, and nonlinear model predictive control of monoclonal antibody (mAb) glycosylation in Chinese hamster ovary (CHO) cell culture (Ma et al., 2026, 2025). This multiscale bioprocess involves coupled submodels and computationally demanding simulations, making it a representative testbed for the capabilities targeted in GlycoPy. In particular, it highlights the need for hierarchical model organization and customized differentiable simulation strategies that can be em-bedded efficiently within gradient-based optimization and control. For this reason, the glycosylation problem serves in this work as a demanding application example for demonstrating the framework.

The remainder of this article is organized as follows. Section 2 introduces the overall structure of GlycoPy and the main ideas underlying its hierarchical modeling architecture and workflow-oriented design. Section 3 presents the core functionalities of the framework, including hierarchical modeling, simulation, parameter estimation, dynamic optimization, and NMPC. Section 4 demonstrates these capabilities through the glycosylation process in CHO cell culture, a representative multiscale bioprocess that highlights the need for hierarchical model development and customized differentiable simulation. Finally, Section 5 concludes the paper and discusses limitations and future directions.

## 2. The Overall Structure of GlycoPy

GlycoPy is organized into two main parts. The first part focuses on modeling and simulation, providing the basic infrastructure for defining process models, assigning variables and equations, and performing numerical or symbolic simulations. The second part builds on this foundation to support parameter estimation, dynamic optimization, and nonlinear model predictive control (NMPC). This division reflects the intended workflow of the framework: models are first constructed and simulated in a structured manner, and are then embedded into optimization and control tasks.

As shown in Fig. 1, in the modeling and simulation part, the central abstraction is the model object, which contains variables, equations, and, when needed, lower-level submodels. Variables are grouped according to their roles as differential variables, algebraic variables, or parameters, while equations are grouped as differential or algebraic equations. This organization allows GlycoPy to assemble the mathematical objects required by a DAE solver while preserving the structure of the original model. As a result, the framework supports equation-oriented modeling in a form that remains compatible with hierarchical model development.

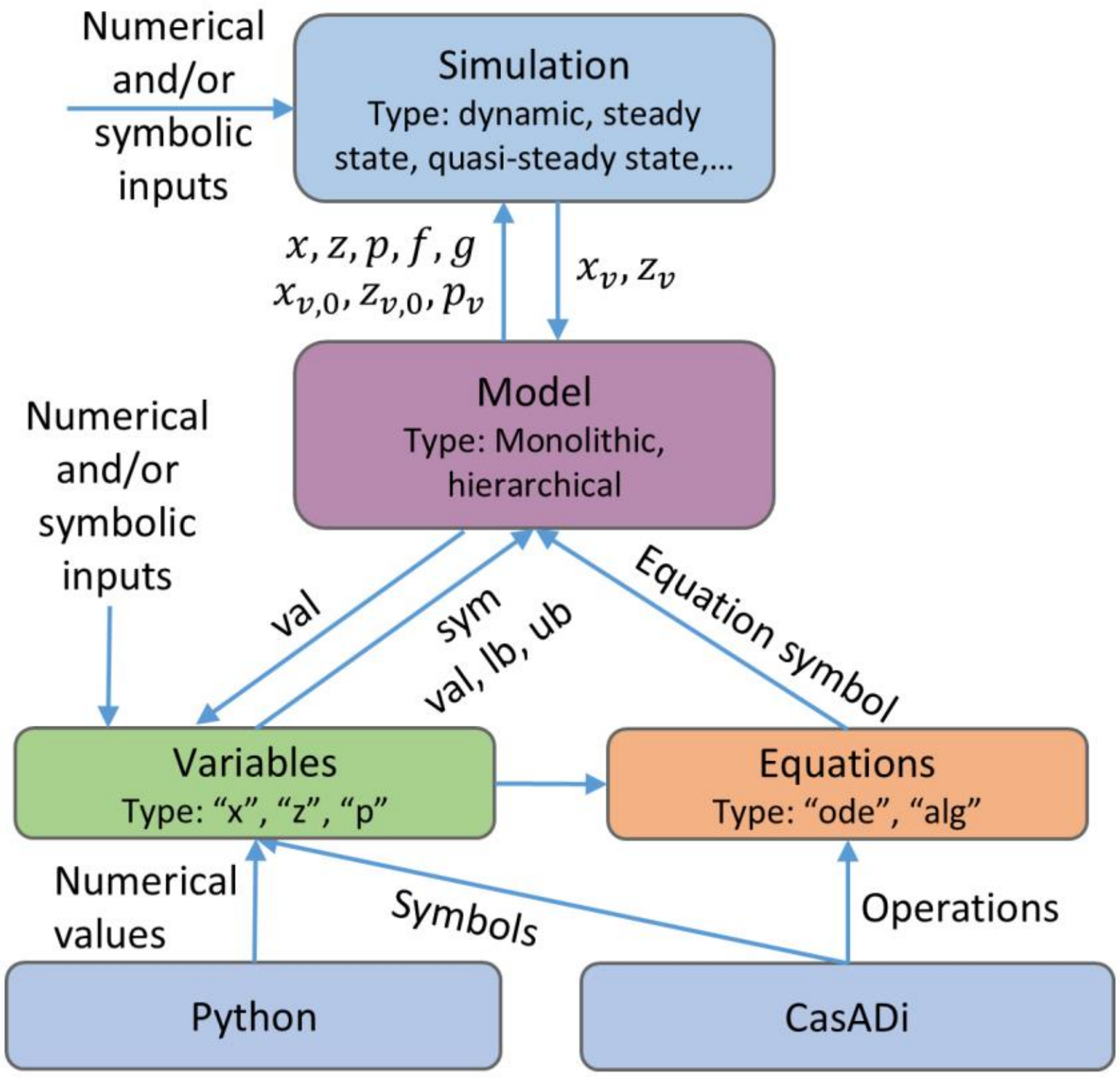

Figure 1. Overall structure of the modeling and simulation layer in GlycoPy. A model in GlycoPy stores variables, equations, and, when needed, lower-level submodels. From these components, the framework assembles the symbolic expressions required for DAE simulation and maps simulation results back to individual variables. Variable symbols, values, and bounds are managed within the model structure, while the simulation layer supports both numerical and symbolic workflows. In the figure, "sym", "val", "lb", and "ub" denote symbol, value, lower bound, and upper bound, respectively; $x$, $z$, and $p$ denote differential variables, algebraic variables, and parameters; and $f$ and $g$ denote differential and algebraic equations.

This organization provides several practical advantages over working directly with low-level symbolic objects alone. First, it supports hierarchical modeling, which is important for the development, maintenance, and reuse of large-scale and multiscale process models. Second, it enables simulation results to be mapped back to individual variables and submodels, which makes the outputs easier to interpret and use in later estimation, optimization, and control tasks. Third, users can assign values directly to selected variables, while GlycoPy automatically assembles the corresponding vectors required by the underlying simulation routines. In this way, the framework adds a model-management layer on top of CasADi's computational capabilities.

A further feature of this layer is that GlycoPy supports both numerical and symbolic simulation. When all inputs are numerical, the simulation returns numerical results in the usual way. When some inputs are symbolic, the simulation returns symbolic expressions that depend on those inputs. This symbolic-simulation capability is important because it allows automatic differentiation to be applied to simulation outputs and, more importantly, enables the construction of differentiable composite simulations. Such workflows are difficult to organize in a reusable way for large hierarchical models when working directly with native CasADi, but they are essential when customized simulation algorithms need to be embedded efficiently within optimization and control.

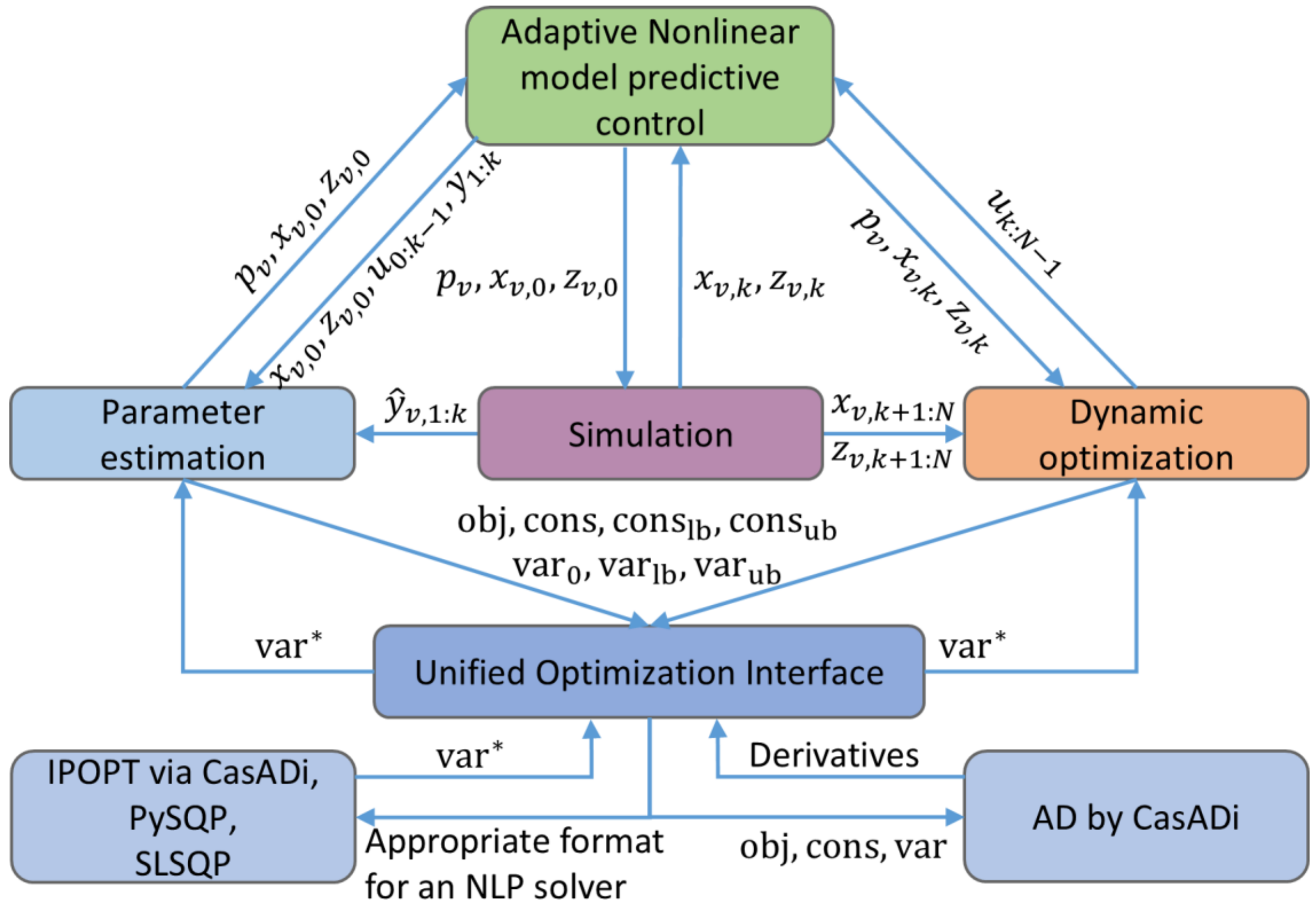

Figure 2. Structure of the estimation, optimization, and control layer in GlycoPy. The framework builds parameter estimation, dynamic optimization, and adaptive nonlinear model predictive control (ANMPC) on top of the same modeling and simulation abstractions. Simulation provides differentiable model predictions for both estimation and optimization, while a unified optimization interface enables different NLP solvers to be used consistently. In the figure, "obj", "cons", "var", and "AD" denote objective function, constraint, variable, and automatic differentiation, respectively.

The second part of GlycoPy uses the above modeling and simulation infrastructure to support parameter estimation, dynamic optimization, and NMPC, as illustrated in Fig. 2. In this part, simulation is used not only for forward prediction but also as a differentiable building block in optimization-based workflows. Parameter estimation combines measurements with model predictions to update unknown parameters and, when necessary, unknown initial conditions. Dynamic optimization computes future control actions over a prediction horizon. NMPC is then constructed by combining repeated estimation, state updating, and dynamic optimization within a closed-loop framework. In this way, the optimization and control components of GlycoPy are built systematically on top of the same model and simulation abstractions.

To support these workflows, GlycoPy also provides a unified optimization interface through which different nonlinear programming solvers can be called in a consistent manner. This design allows the same estimation or control formulation to be used with different solvers while retaining access to derivative information generated through CasADi. Based on these components, users can implement both state NMPC and adaptive NMPC (ANMPC) within a common framework. The next section introduces the main functionalities of GlycoPy in more detail, including hierarchical modeling, simulation, parameter estimation, dynamic optimization, and NMPC.

## 3. Core Functionalities of GlycoPy

This section introduces the core functionalities of GlycoPy, with emphasis on the capabilities most relevant to large-scale and multiscale bioprocess applications. We first describe hierarchical modeling and simulation, which form the foundation of the framework, and then summarize how these components are used in parameter estimation, dynamic optimization, and NMPC.

### 3.1 Hierarchical modeling

In GlycoPy, a model is defined as an object that stores variables, equations, and, when needed, submodels, all of which are themselves objects. Variables in GlycoPy can be of type $p$ (parameter), $x$ (differential variable), or $z$ (algebraic variable), while equations can be of type $f$ (differential equation) or $g$ (algebraic equation). Based on this information, the model can assemble the symbolic expressions required for DAE simulation and can also map vectorized simulation results back to individual variables. This design allows users to develop process models in an equation-oriented manner while maintaining direct access to variable values and trajectories for later use in simulation, estimation, optimization, and control. In particular, because variables are stored as objects within the model, their current values and full trajectories can be accessed directly when needed.

The model abstraction in GlycoPy is implemented in an object-oriented manner, which makes it natural to construct related models through inheritance and to build more complex models through composition. As a result, a GlycoPy model may contain other models as submodels, thereby forming a hierarchical structure, as illustrated in Fig. 3. This hierarchical organization is particularly useful for large-scale and multiscale processes, for which different physical or biological subsystems are more naturally developed and tested as separate components before being integrated into a full process model.

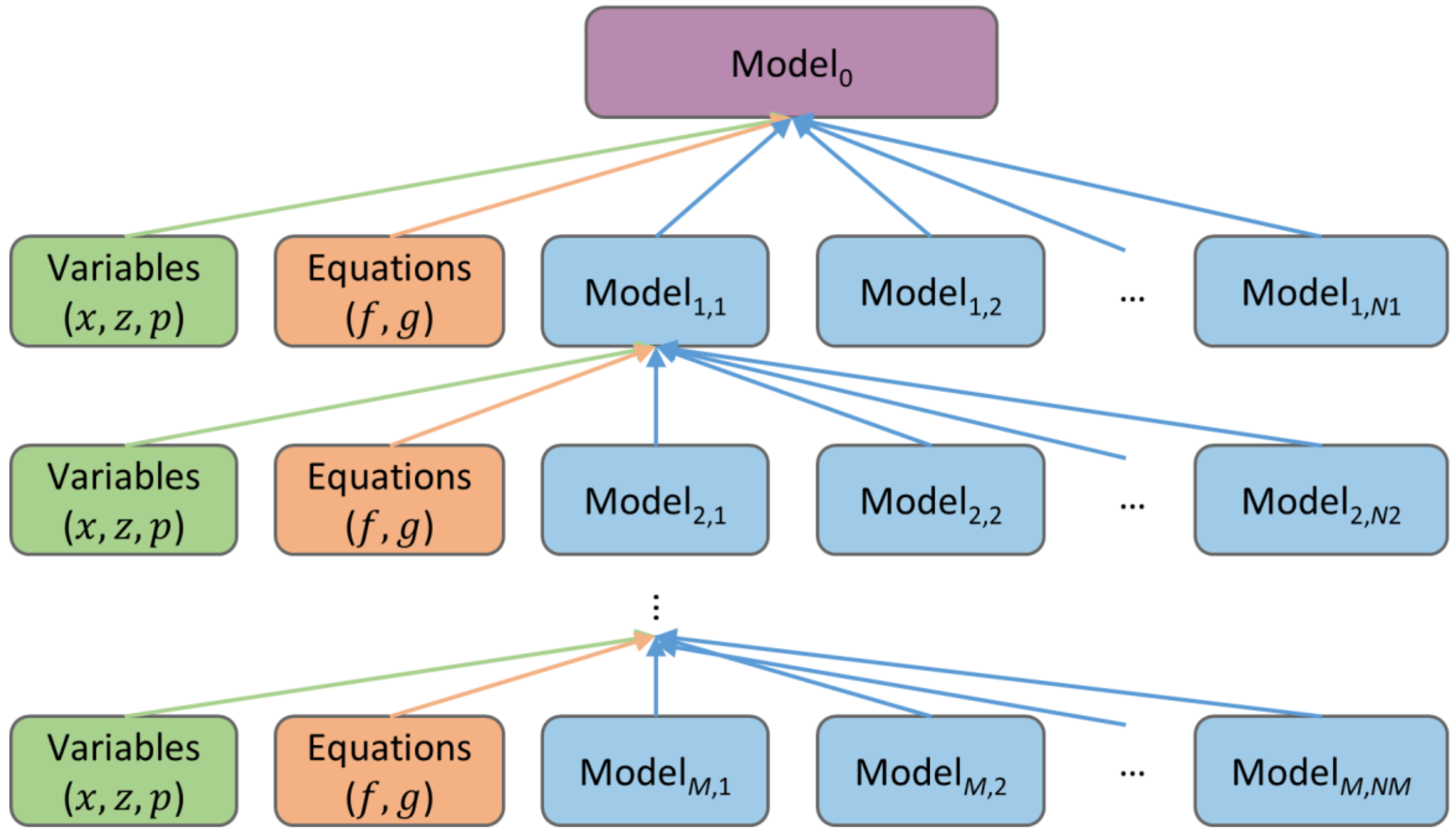

Figure 3. Schematic illustration of a hierarchical model in GlycoPy. Each model or submodel is represented as an object containing variables and equations, and may itself include lower-level submodels. Model$_{n,m}$ refers to the $n$-th submodel in the $m$-th layer of the hierarchical model.

In a hierarchical model, lower-level submodels are designed to remain relatively independent of the upper-level model, which facilitates model development, debugging, and reuse. During model assembly, the variables and equations defined in the submodels are combined with those defined at the current level of the hierarchy, as illustrated in Fig. 4. This combination proceeds from the lowest level upward, so that the top-level model ultimately contains all variables and equations required by the DAE solver. In this way, GlycoPy preserves the modular structure of the original model while still producing a unified mathematical representation for simulation and optimization.

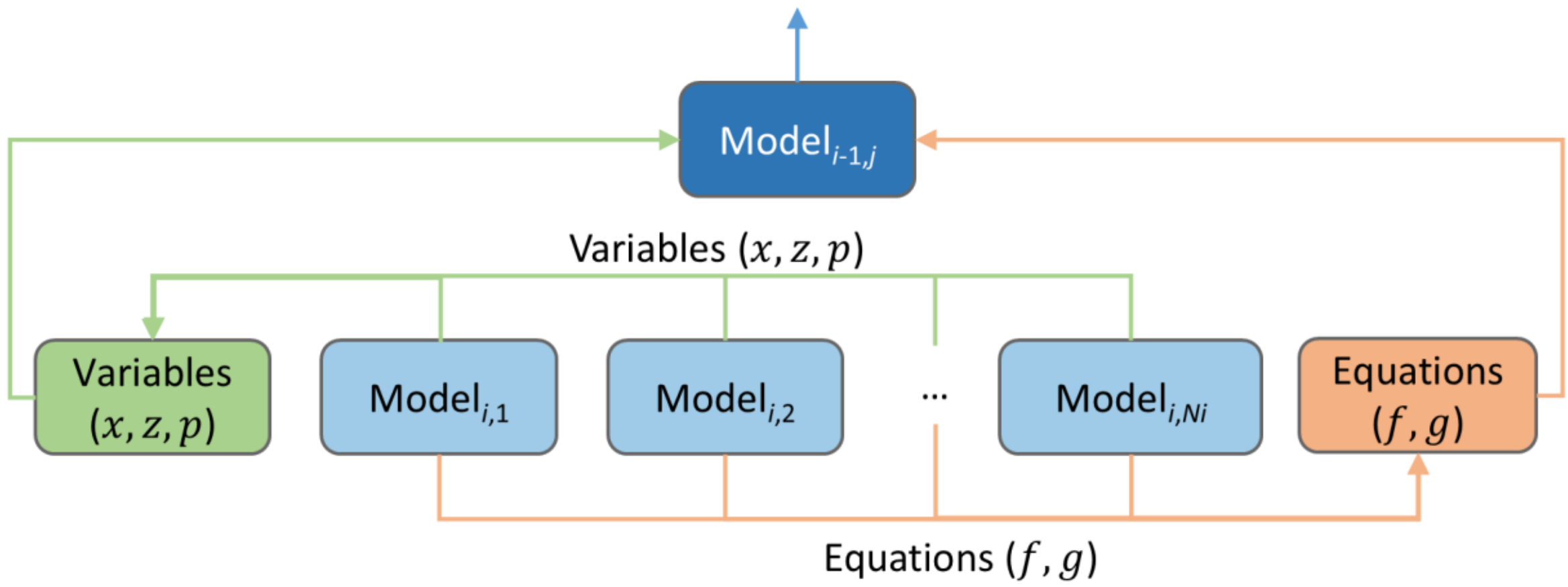

Figure 4. Assembly of a hierarchical model in GlycoPy. Variables and equations from lower-level submodels are combined level by level into the upper-level model, so that the top-level model contains the complete set of equations required for simulation.

This hierarchical modeling strategy improves the adaptability, maintainability, and usability of complex process models. Instead of building a large model as a single monolithic system, users can organize the model around physically meaningful subsystems and their interactions. This is particularly important for chemical and bioprocess applications, where process models often involve many coupled units, multiple scales, and heterogeneous state and parameter sets. The hierarchical structure of GlycoPy therefore provides an important foundation for the customized simulation, optimization, and NMPC workflows described in the following sections.

### 3.2 Simulation

Simulation in GlycoPy is built on the DAE solver IDAS in SUNDIALS (Gardner et al., 2022) through CasADi. The framework considers models of the form

$$\frac{dx}{dt} = f(x, z, p) \tag{1}$$

$$0 = g(x, z, p) \tag{2}$$

where $x \in \mathbb{R}^{N_x}$, $z \in \mathbb{R}^{N_z}$, and $p \in \mathbb{R}^{N_p}$ denote differential variables, algebraic variables, and parameters, respectively. The functions $f$ and $g$ are assembled automatically from the GlycoPy model and passed to the underlying DAE solver for simulation.

In addition to basic DAE simulation, GlycoPy provides functionality for consistent initialization. Although a DAE solver typically requires initial values for the differential variables, in practice it may be more convenient to specify some algebraic variables directly. To handle such situations, GlycoPy determines consistent initial values by solving the algebraic equations together with the unspecified initial conditions before time integration begins. This feature makes the simulation workflow more flexible and more convenient for process applications. For example, in the mass balance equations of a bioreactor, the mole amounts of species $i$, denoted by $d_i$, may be chosen as the differential variables $x$, whereas the molar concentrations $c_i$, appearing as algebraic variables $z$, are often more natural for users to initialize. Given the initial values of $c_i$, the corresponding values of $d_i$ can be computed automatically during the initialization by solving the algebraic equations $d_i = Vc_i$, where $V$ is the bioreactor volume.

GlycoPy also supports simulations with time events. In many process applications, some parameter values, such as feed flow rates, are piecewise constant and change only at predefined time points. In GlycoPy, these time-varying inputs are handled by restarting the DAE solver at each event with updated parameter values and initial conditions taken from the previous simulation segment. This strategy provides a transparent and flexible way to manage event-driven simulations, while also making it easier to inspect intermediate results and diagnose the behavior of the model. The overall workflow is illustrated in Fig. 5.

Figure 5. Simulation workflow in GlycoPy. The framework assembles variables and equations from the model and performs DAE simulation with numerical or symbolic inputs. For time-event simulations, the solver is restarted at predefined event times with updated parameter values and initial conditions carried over from the previous segment. Numerical inputs lead to numerical trajectories, whereas symbolic inputs produce symbolic outputs that can be differentiated or reused in later simulations, thereby enabling differentiable composite simulation workflows.

A key feature of the simulation module is that it supports both numerical and symbolic simulation. When all inputs are numerical, the simulation returns numerical trajectories in the usual way. When some inputs are symbolic, the simulation returns symbolic expressions that depend on those inputs. These symbolic outputs can be treated like other CasADi symbolic expressions: they can be differentiated automatically, used to define functions, or passed as inputs to subsequent simulation steps. As a result, GlycoPy supports not only ordinary forward simulation but also differentiable composite simulation workflows.

This symbolic-simulation capability is particularly important for customized simulation algorithms that need to be embedded within optimization and control. When only a few simulations are required, direct numerical simulation is often more convenient. However, when many repeated simulations and derivative evaluations are needed, symbolic simulation becomes especially useful. Such workflows are difficult to develop and reuse for large hierarchical models when working directly with native CasADi, but they are essential when customized differentiable simulation algorithms must be embedded efficiently within optimization and control.

**3.3 Dynamic optimization**

Dynamic optimization in GlycoPy is formulated as an optimal control problem in which the time-varying control input is selected to optimize a prescribed performance objective subject to the process dynamics, algebraic constraints, and input/state constraints. In general, the problem can be written as

$$\min_{x(\cdot),z(\cdot),u(\cdot)} \int_0^T L(x(t), z(t), u(t), p)dt + E(x(T), z(T), p) \qquad \text{(OCP)}$$

subject to the DAE model, the initial and terminal conditions, and the admissible sets for the states and control variables. Here, $L(\cdot)$ and $E(\cdot)$ denote stage and terminal cost functions, respectively. The variables $x(t)$, $z(t)$, and $u(t)$ represent the differential states, algebraic states, and control inputs, respectively, and $p$ denotes the model parameters.

To solve this problem numerically, GlycoPy transcribes it into a nonlinear programming (NLP) problem. In principle, dynamic optimization problems can be transcribed by several direct methods, such as control vector parameterization (CVP), multiple shooting, and collocation. In the current implementation, GlycoPy adopts CVP because it aligns naturally with the simulation-oriented structure of the framework: the control trajectory is parameterized over a finite number of intervals, and the resulting simulations are used to evaluate the objective function and constraints. This design allows the dynamic optimization module to make direct use of the differentiable simulation capabilities described in Section 3.2.

Based on this formulation, GlycoPy constructs the corresponding NLP and solves it using the unified optimization interface introduced in Section 2. The framework therefore links model development, simulation, and optimization in a consistent workflow, so that the same hierarchical process model can be used directly in dynamic optimization and, later, in NMPC. In this way, dynamic optimization in GlycoPy serves both as a standalone capability and as a core building block for the control applications considered in this work.

### 3.4 Parameter estimation

Parameter estimation in GlycoPy is formulated as an optimization problem in which model parameters and, when necessary, unknown initial conditions are adjusted to improve agreement between model predictions and experimental measurements. In general, this problem is constrained by the same DAE model used for simulation, together with the measured inputs and any path or terminal constraints imposed on the states or outputs. Thus, parameter estimation in GlycoPy is built directly on the modeling and simulation framework described above, rather than being treated as a separate workflow.

As in dynamic optimization, the parameter-estimation problem is transcribed into a nonlinear programming (NLP) problem and solved through the unified optimization interface. A practical difference is that a single objective evaluation may involve multiple simulations, since data from several experiments can be fitted simultaneously. GlycoPy organizes these simulations and residual calculations automatically, thereby making multi-experiment parameter estimation more systematic and reusable for complex hierarchical models.

GlycoPy currently supports several standard estimators, including ordinary least squares (OLS), maximum likelihood (ML), and maximum a posteriori (MAP) estimation. In addition to the estimated parameter values, the framework can also return covariance information based on the sensitivities of the model outputs with respect to the estimated parameters. These capabilities make the parameter-estimation module suitable both for offline model fitting and for the repeated parameter updates required in adaptive NMPC.

### 3.5 Nonlinear model predictive control

By combining parameter estimation, simulation, and dynamic optimization, GlycoPy supports nonlinear model predictive control (NMPC) with or without online parameter adaptation, referred to here as adaptive NMPC (ANMPC) and state NMPC, respectively. In both cases, the controller repeatedly uses the process model to predict future behavior over a finite horizon and computes control actions by solving a dynamic optimization problem. In ANMPC, the model parameters are additionally updated online from newly available measurements, allowing the controller to adapt to plant-model mismatch and process variations. In the present paper, the detailed workflow is illustrated using ANMPC, since it is the more general case; state NMPC is supported as the simpler special case without online parameter adaptation.

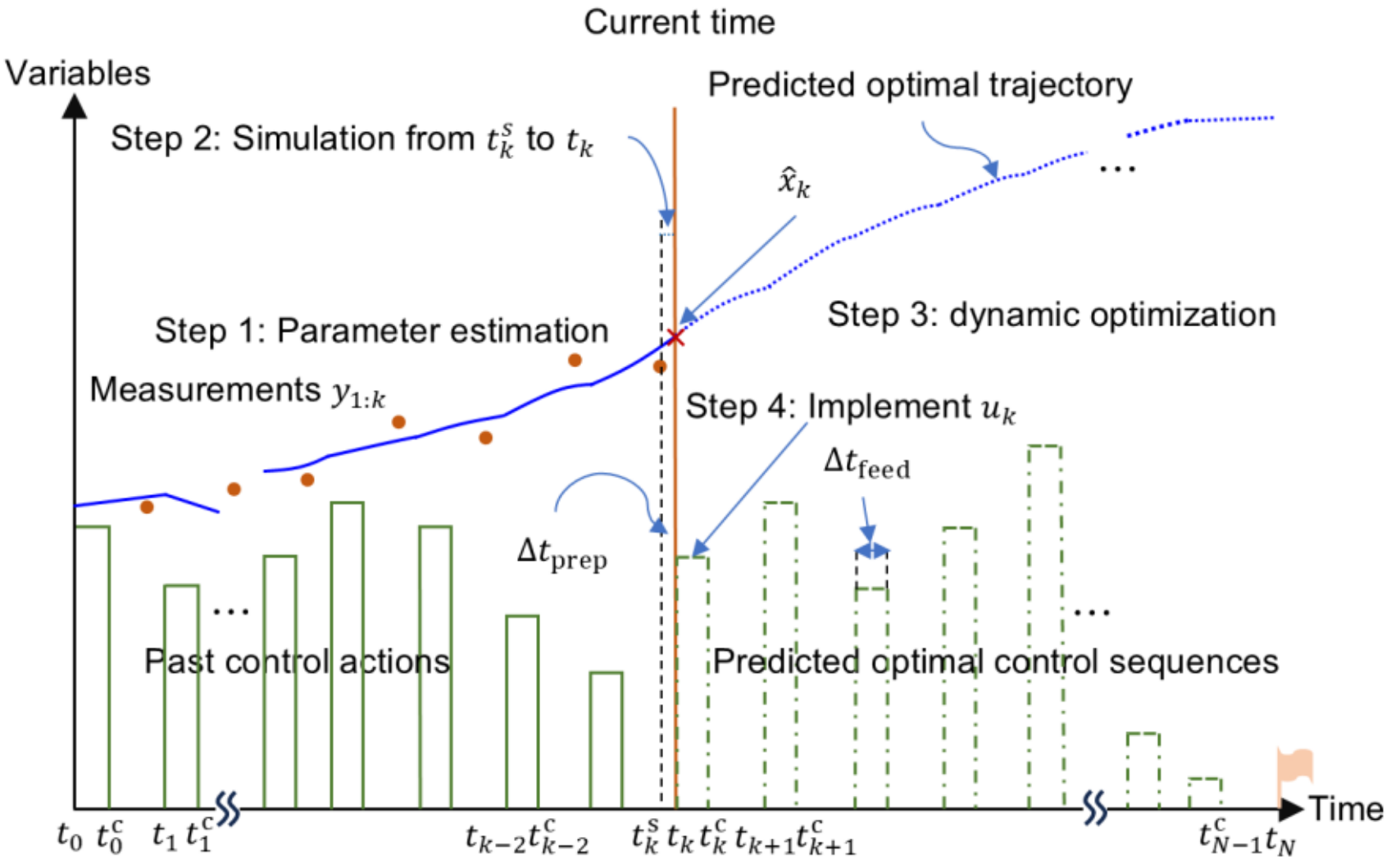

Figure 6. Adaptive nonlinear model predictive control (ANMPC) for a fed-batch process in a shrinking-horizon framework. At each control step, available measurements are used to update the model parameters (Step 1), the process model is simulated to estimate the current state (Step 2), and a dynamic optimization problem is solved over the remaining horizon to compute the future control sequence (Step 3). Only the first control action is applied to the process (Step 4) before the procedure is repeated at the next sampling time. $\Delta t_{\mathrm{prep}}$ is the preparation time; $\Delta t_{\mathrm{feed}}$ is the feeding time.

For fed-batch processes, GlycoPy implements NMPC in a shrinking-horizon form, as illustrated in Fig. 6. At each control step, the available measurements are first used for parameter updating when adaptation is enabled (Step 1). The model is then simulated to estimate the current process state (Step 2), after which a dynamic optimization problem is solved over the remaining horizon to compute the future control sequence (Step 3). Only the first control action is applied to the process (Step 4), and the procedure is repeated when new measurements become available. In this way, NMPC in GlycoPy is constructed systematically from the same simulation, estimation, and optimization modules introduced in the previous sections.

In addition to the control logic itself, GlycoPy is designed to support practical implementation and debugging of NMPC workflows. Intermediate results, such as updated parameters, estimated states, simulation trajectories, and optimized control actions, can be stored and accessed step by step, which makes it easier to inspect, restart, and analyze the controller during operation. As illustrated in Fig. 7, this modular design allows users to tailor the estimation, state-update, and optimization steps to their specific application while retaining a consistent workflow for closed-loop control.

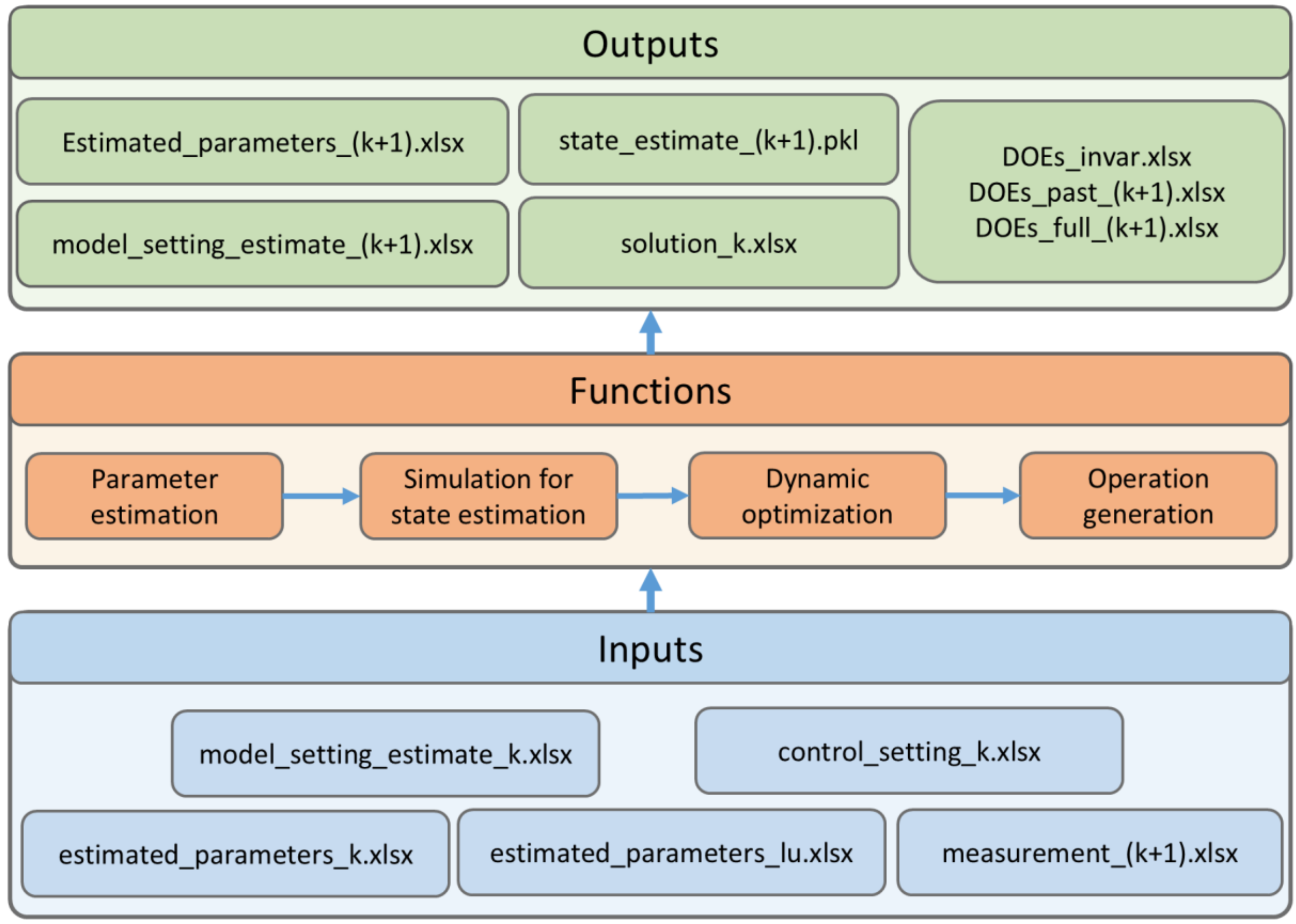

Figure 7. Workflow for one ANMPC step in GlycoPy. The framework combines parameter estimation, state updating, and dynamic optimization in a modular manner, with intermediate results such as updated parameters, estimated states, predicted trajectories, and control actions stored for later inspection and reuse. This design facilitates implementation, debugging, and stepwise restart of closed-loop control workflows.

## 4. Demonstration on a Multiscale Glycosylation Process

This section demonstrates GlycoPy on the multiscale glycosylation process in CHO cell culture, a representative example chosen because it simultaneously challenges the framework in model organization, simulation efficiency, and control implementation. The process couples cell culture dynamics, intracellular nucleotide sugar donor synthesis, and Golgi glycosylation reactions, resulting in a large-scale model that is difficult to handle directly in NMPC. It therefore provides a stringent test of the main ideas underlying GlycoPy. Specifically, Section 4.1 illustrates the hierarchical construction of the multiscale model, Section 4.2 shows how symbolic simulation enables a customized differentiable quasi-steady-state (QSS) simulation strategy, and Section 4.3 demonstrates the integrated ANMPC workflow for glycosylation control.

### 4.1 Hierarchical modeling of the multiscale glycosylation process

The multiscale glycosylation process considered in this work is illustrated in Fig. 8. The model is based on the cell culture and metabolism model, the intracellular nucleotide sugar donor (NSD) synthesis model, and the Golgi glycosylation model, which are coupled through extracellular metabolites, intracellular precursors, and glycan distributions.

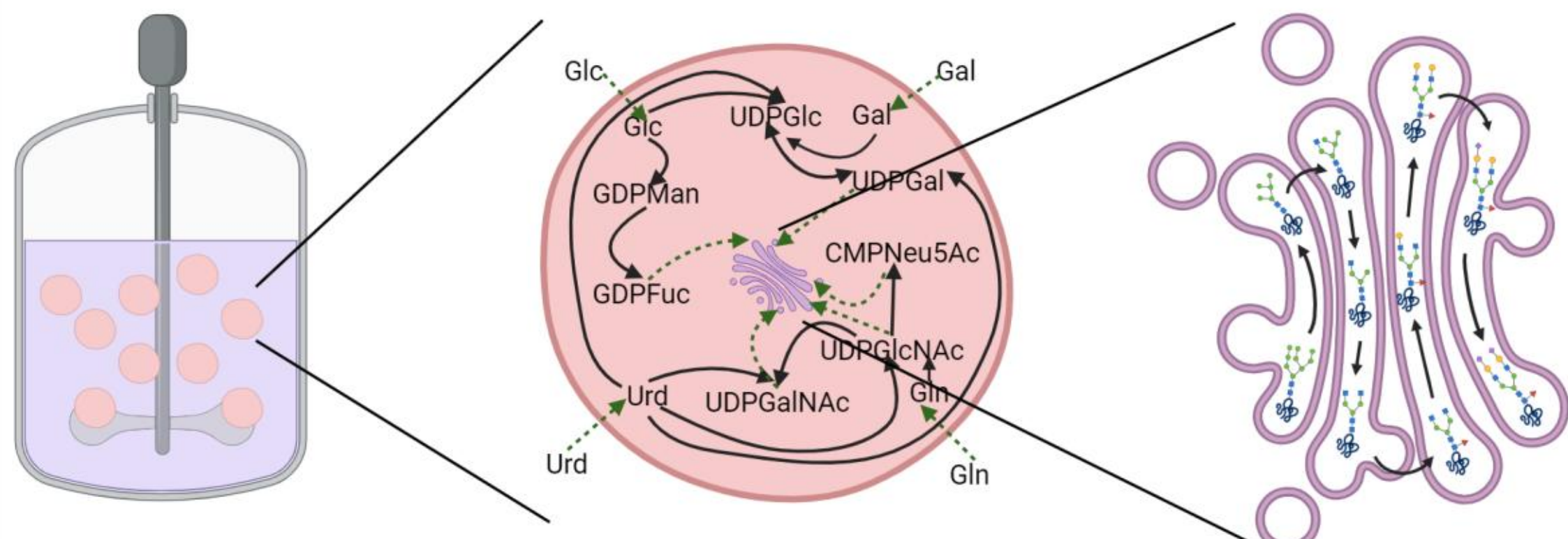

Figure 8. Multiscale mAb glycosylation process considered in this work. The overall process is decomposed into three coupled parts: the cell culture and metabolism submodel, the intracellular nucleotide sugar donor (NSD) synthesis submodel, and the Golgi glycosylation submodel. Together, these submodels describe how extracellular culture conditions, intracellular precursor synthesis, and Golgi reaction dynamics jointly determine the glycan profile of the produced mAb. Reprinted from (Ma et al., 2025) with permission.

Implementing this process as a single monolithic model would be inconvenient for development, testing, maintenance, and later use in simulation, parameter estimation, optimization, and control. For this reason, the model is organized hierarchically in GlycoPy, as shown in Fig. 9. This example serves to demonstrate how the hierarchical modeling strategy introduced in Section 3.1 can be used to construct a large and strongly coupled bioprocess model in a structured and reusable manner.

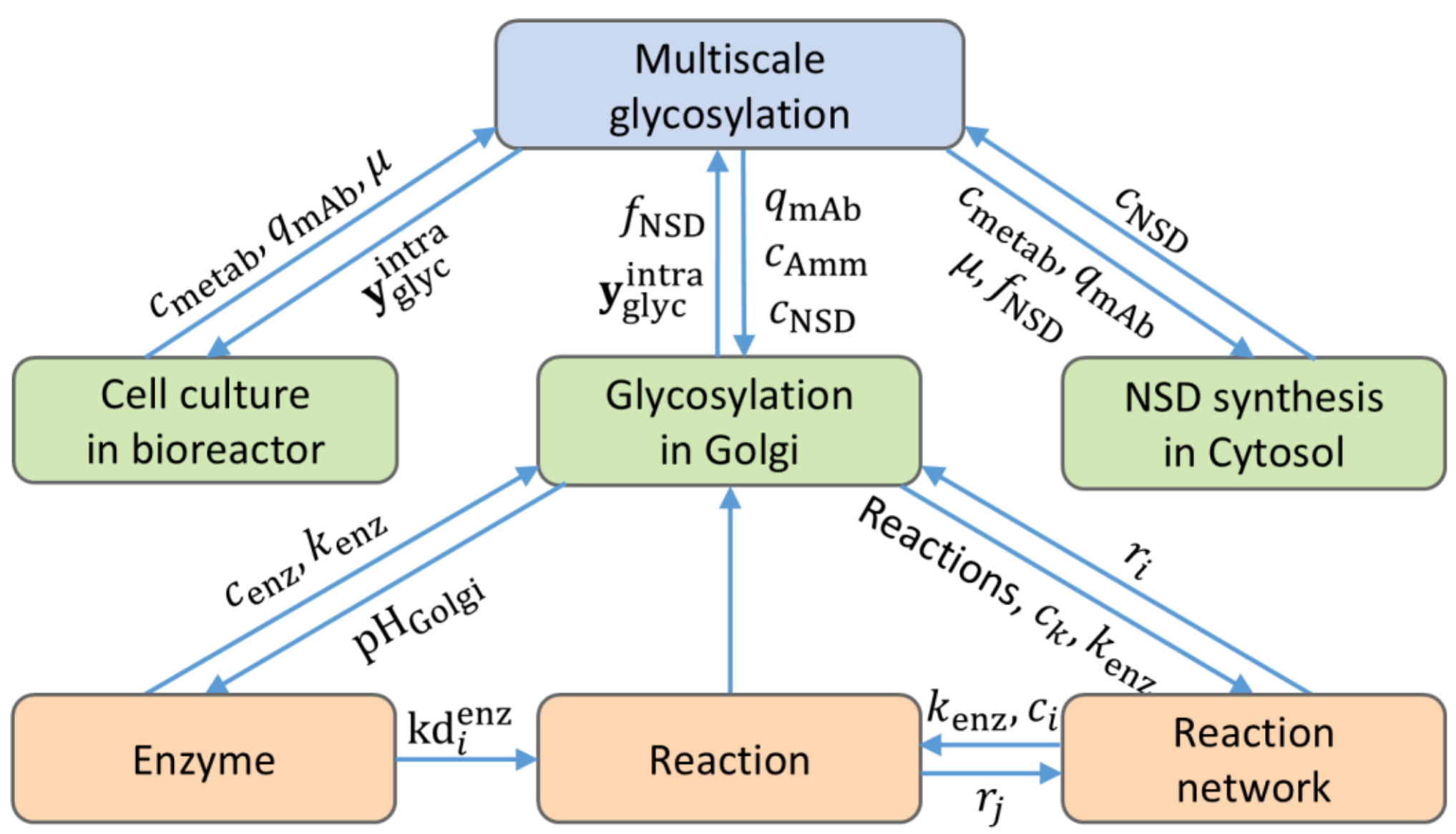

Figure 9. Hierarchical implementation of the multiscale glycosylation model in GlycoPy. The top-level model coordinates information exchange among the cell culture, NSD synthesis, and Golgi glycosylation submodels, while the Golgi submodel is further organized into lower-level components. Arrows indicate the main information flows among submodels, illustrating how GlycoPy preserves the subsystem structure of the process while assembling a unified model for simulation and control. Notation**:** $c_{\text{Amm}}$, $c_{\text{enz}}$, $c_i$, $c_k$, $c_{\text{metab}}$, and $c_{\text{NSD}}$ denote concentrations; $f_{\text{NSD}}$ denotes the NSD consumption rate; $k_{\text{enz}}$ and $k_{d,i}$ denote enzyme kinetic and dissociation constants; $\text{pH}_{\text{Golgi}}$ denotes Golgi pH; $q_{\text{mAb}}$ denotes the specific mAb production rate; $r_i$ and $r_j$ denote component and reaction rates, respectively; Reactions denotes a list of Reaction submodels; $\mathbf{y}_{\text{glyc}}^{\text{intra}}$ denotes the intracellular glycan percentage; and $\mu$ denotes the cell growth rate.

At the highest level, the multiscale model is decomposed into three main submodels: the cell culture model, the NSD synthesis model, and the Golgi glycosylation model, while a main model coordinates the information exchange among them. The cell culture submodel provides extracellular metabolite concentrations ($c_{\text{metab}}$), cell growth information ($\mu$), and production-related quantities ($q_{\text{mAb}}$) required by the other submodels. It also receives intracellular glycan information from the Golgi model to compute extracellular glycan distributions. The NSD synthesis submodel computes intracellular nucleotide sugar donor concentrations ($c_{\mathrm{NSD}}$) using information from both the cell culture model and the Golgi model, including the NSD consumption rate in the Golgi ($f_{\mathrm{NSD}}$). In this way, it links cellular metabolism with glycosylation dynamics. This decomposition preserves the physical structure of the process while making the interactions among different scales explicit.

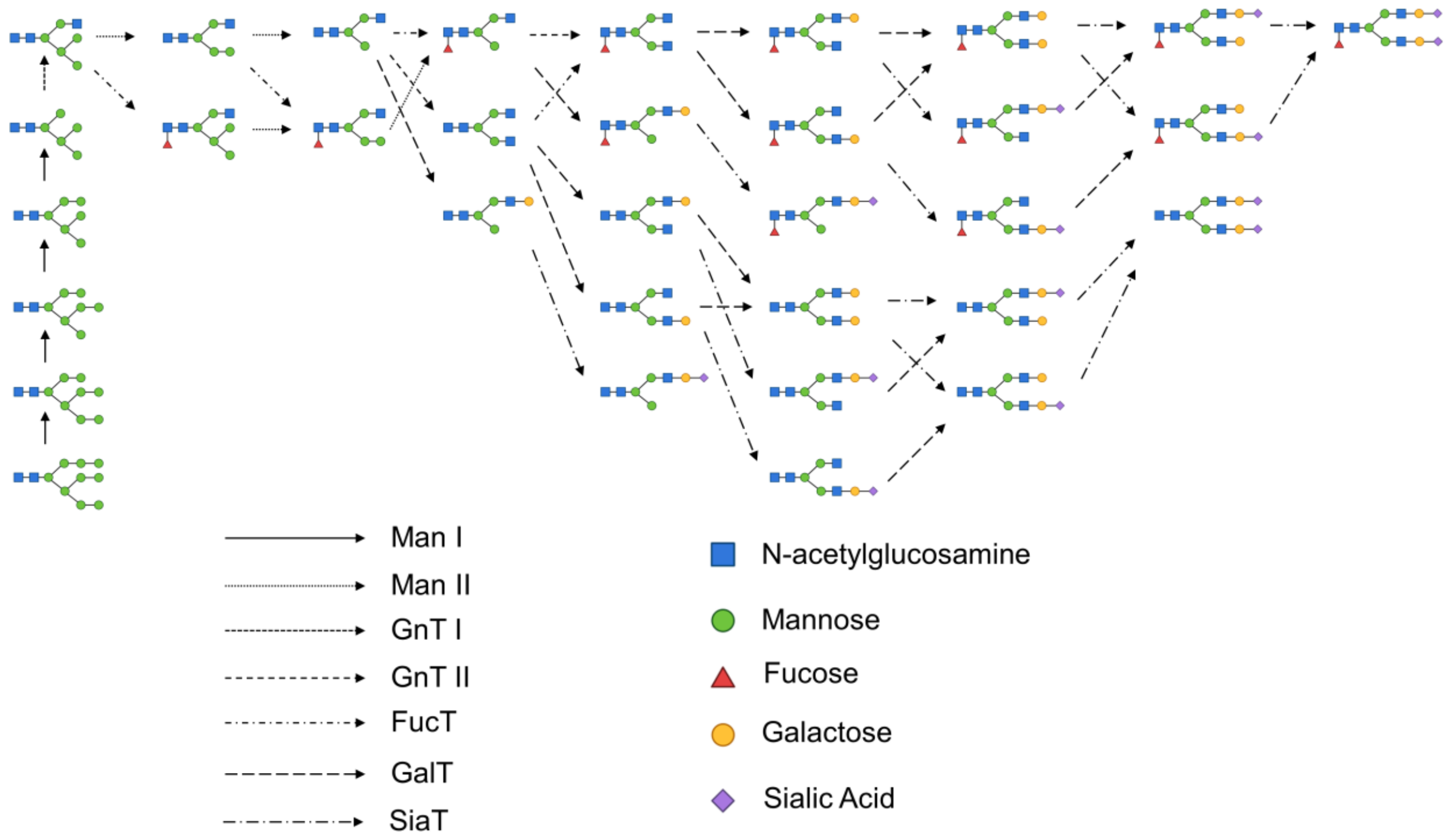

Figure 10. Glycosylation reaction network in the Golgi submodel. The network comprises 43 reactions involving 33 oligosaccharides and 7 enzymes, and represents the most complex part of the multiscale model. Its size and kinetic complexity motivate the use of hierarchical modeling in GlycoPy. Reprinted from (Ma et al., 2026) with permission.

Because the Golgi glycosylation submodel contains the most complex reaction network (Fig. 10) and kinetic relations, it is further decomposed into lower-level submodels, including enzyme, reaction, and reaction-network submodels. The Golgi main model describes the mass balances of oligosaccharides and NSDs in the Golgi apparatus and manages the information exchange among the submodels to compute the corresponding concentrations ($c_i$, $c_k$). The lower-level submodels, in turn, provide enzyme properties, reaction-specific rates $\left(r_j\right)$, and aggregated network-level rates ($r_i$). This additional layer of hierarchy allows the large number of reactions, enzymes, and associated kinetic relations to be organized in a modular way, rather than embedding all expressions directly into a single model block.

This case study highlights the practical value of hierarchical modeling in GlycoPy. By organizing the multiscale glycosylation process around physically meaningful subsystems and their interactions, the framework makes the model easier to develop, test, debug, maintain, and reuse. At the same time, all variables and equations from the main model and submodels can still be assembled into a unified mathematical representation for simulation and control. Compared with working directly with native CasADi, GlycoPy provides a more structured and reusable way to implement this hierarchical model, reducing the manual bookkeeping required across submodels and making the overall model more manageable for large-scale and multiscale applications.

### 4.2 Customized differentiable quasi-steady-state simulation

Direct simulation of the full multiscale glycosylation model is computationally expensive, which makes its direct use in optimization and NMPC impractical. To address this challenge, a QSS simulation strategy is adopted for the present application. This example is particularly important for GlycoPy because it demonstrates not only the need for customized simulation algorithms in multiscale bioprocesses, but also the ability of the framework to implement such algorithms in a differentiable and reusable manner.

The QSS simulation strategy can be summarized as follows. First, the cell culture and NSD submodels are simulated together at a series of selected time points, and the resulting environment variables at each time point, denoted by $\mathbf{y}_{\mathrm{env}}^{n}$, are used to define the conditions for the Golgi glycosylation submodel. Second, at each selected time point, the Golgi submodel is simulated independently under a quasi-steady-state assumption to obtain the corresponding intracellular glycosylation profile ($\mathbf{y}_{\mathrm{glyc}}^{\mathrm{intra},1}$). Finally, the cell culture model is simulated again by treating these intracellular glycosylation quantities as time-varying inputs, thereby recovering the extracellular glycan percentage ($\mathbf{y}_{\text{glyc}}^{\text{extra},1}$) trajectories. Because the Golgi simulations at different time points are independent, they can be executed in parallel. More details of the algorithm can be found in (Ma et al., 2025).

In GlycoPy, this QSS strategy is implemented as a customized simulation workflow built on top of the symbolic simulation capability described in Section 3.2, as illustrated in Fig. 11. The symbolic results generated at the selected time points are reused as inputs to the later steady-state Golgi simulations, and the resulting intracellular glycosylation profiles are then reused again in the subsequent cell culture simulation. In this way, the overall QSS procedure is constructed as a differentiable composite simulation rather than as a disconnected sequence of numerical calculations. This is important because the resulting workflow remains compatible with gradient-based optimization and control.

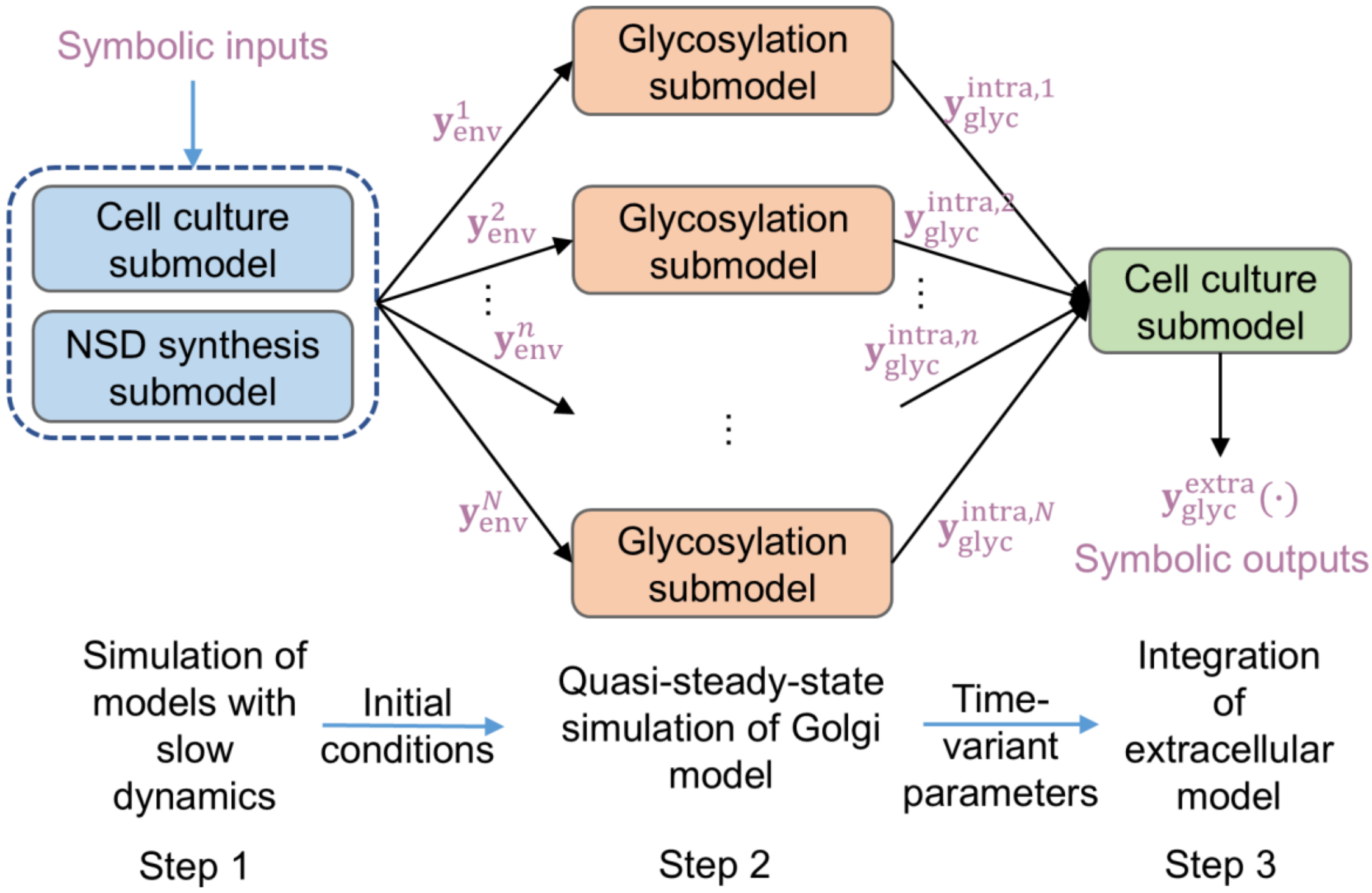

Figure 11. Symbolic and differentiable QSS simulation workflow in GlycoPy. The cell culture and NSD submodels are first simulated at selected time points to generate symbolic environment variables for the Golgi submodel. Independent steady-state Golgi simulations are then carried out at these time points to obtain intracellular glycosylation profiles, which are subsequently reused as time-varying inputs in the final cell culture simulation. This composite workflow illustrates how GlycoPy supports customized differentiable simulation strategies for multiscale models.

This example also highlights the practical value of GlycoPy relative to working directly with low-level and fundamental symbolic tools. Although a similar QSS algorithm could in principle be assembled directly using native CasADi, doing so for a large hierarchical multiscale model would require substantial manual organization of variables, equations, intermediate symbolic quantities, and data flow across submodels. In GlycoPy, the hierarchical model structure and symbolic simulation interface make it much easier to construct, reuse, and embed such customized differentiable simulation workflows within optimization and control.

As reported in our previous work, the parallel QSS strategy accelerates the simulation of the multiscale glycosylation process by more than two orders of magnitude (Ma et al., 2025). More importantly for the present paper, it provides a concrete demonstration of one of the main capabilities of GlycoPy: the framework enables users to move beyond standard forward simulation and to implement specialized differentiable simulation algorithms for challenging multiscale applications.

### 4.3 ANMPC of the glycosylation process

The multiscale glycosylation model and the QSS simulation strategy are next used for nonlinear model predictive control of the CHO cell culture process. Because many model parameters are uncertain and may vary from their nominal values, adaptive nonlinear model predictive control (ANMPC) is adopted so that parameter estimates can be updated online from newly available measurements (Ma et al., 2026). This application serves to demonstrate the full workflow enabled by GlycoPy, namely the integration of hierarchical modeling, customized differentiable simulation, parameter estimation, and dynamic optimization within a closed-loop control framework.

In this case study, the manipulated inputs are the media supplement and galactose supplement. The control objective is to maximize the galactosylation index (GI) at harvest, defined as

$$\mathrm{GI}(T) = [\mathrm{FA2G1}](T) + 2\,[\mathrm{FA2G2}](T), \tag{3}$$

where $[\mathrm{FA2G1}](T)$ and $[\mathrm{FA2G2}](T)$ denote the concentrations of FA2G1- and FA2G2-attached mAb at the final time, respectively. Three constraints are imposed in the control problem: the cell viability must remain no less than 60%, the bioreactor volume must remain between 0.75 L and 2.25 L, and the percentage of Man5-attached mAb must remain no greater than 5% (Ma et al., 2026). These specifications make the control problem representative of practical bioprocess operation, where product quality and process feasibility must be considered simultaneously.

The ANMPC implementation follows the modular workflow described in Section 3.5. At each control step, model parameters are updated from the available measurements, the current process state is estimated through simulation, and a dynamic optimization problem is solved over the remaining horizon to compute the future control actions. For the present multiscale model, parameter updating is performed sequentially for the cell culture submodel and for the remaining submodels in order to reduce overfitting and improve computational efficiency. This application therefore illustrates not only the closed-loop control capability of GlycoPy, but also its flexibility in accommodating problem-specific estimation and control workflows within a consistent framework.

The resulting trajectories are shown in Fig. 12, where ANMPC is compared with ground-truth optimization, state NMPC and open-loop optimization. Here, the ground-truth optimization refers to the optimization solved using the ground-truth parameter values employed to generate the synthetic plant behavior, whereas the other control schemes start from incorrect initial parameter values. As shown in Figs. 12a–c, all three control strategies satisfy the imposed process constraints, indicating that the optimization problems are solved feasibly. However, ANMPC achieves a substantially higher galactosylation index than open-loop optimization and state NMPC, while remaining close to the ground-truth optimum. This improvement arises because online parameter adaptation reduces plant–model mismatch, allowing the controller to maintain feasible operation while steering the process more effectively toward the desired terminal glycosylation outcome. More broadly, this result demonstrates that GlycoPy can translate a large hierarchical model and a customized differentiable simulation strategy into an effective ANMPC implementation for multiscale bioprocess control.

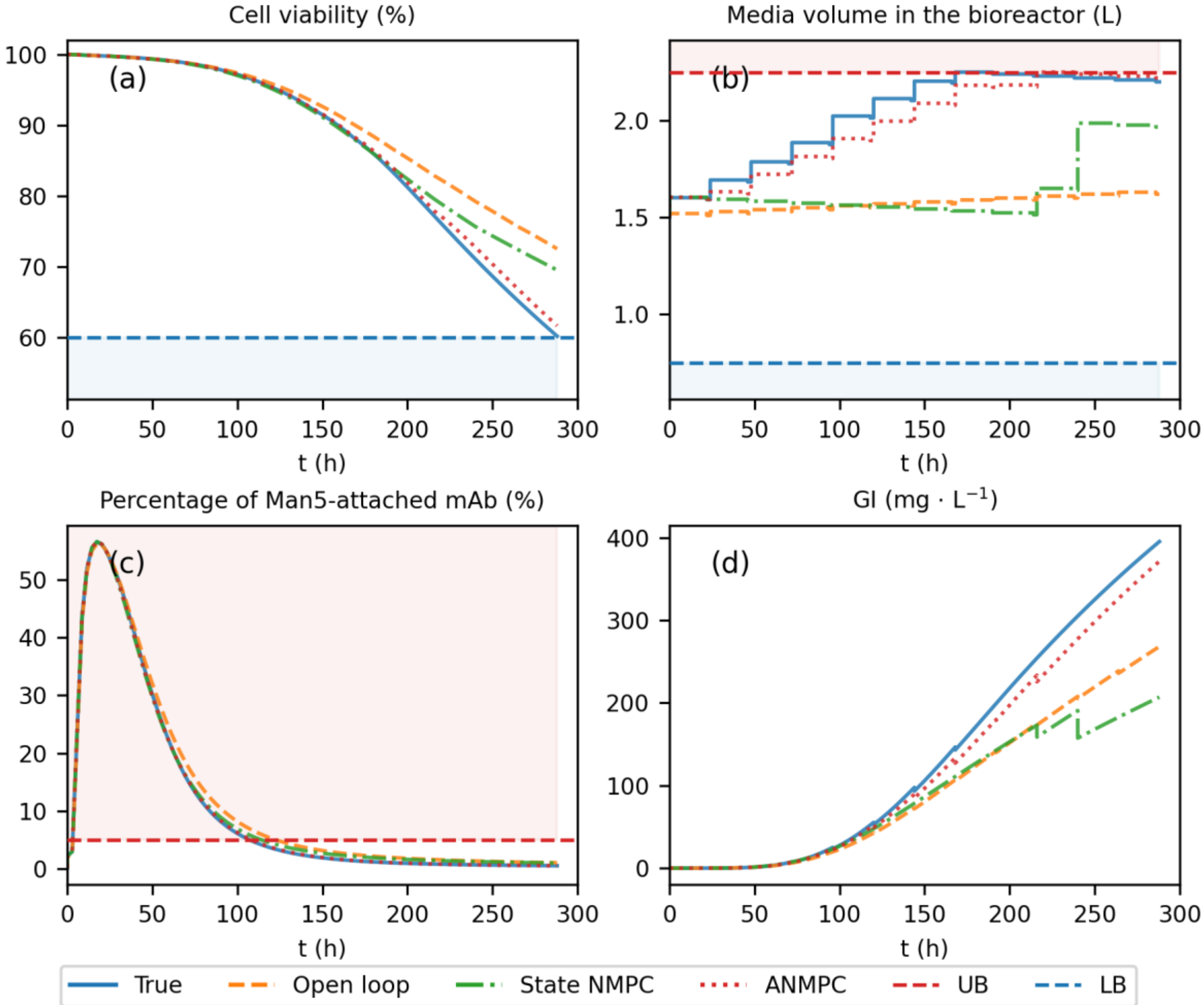

Figure 12. Comparison of the closed-loop results obtained by ground-truth optimization, open-loop optimization, state NMPC, and ANMPC for the glycosylation process in CHO cell culture: (a) cell viability, (b) bioreactor volume, (c) percentage of Man5-attached mAb, and (d) galactosylation index (GI). LB and UB denote the lower and upper constraint bounds, respectively. All control strategies satisfy the imposed process constraints, whereas ANMPC achieves the highest GI among the implementable control schemes and remains close to the ground-truth optimum. The results illustrate the benefit of online parameter adaptation for improving product quality while maintaining feasible process operation.

## 5. Conclusions

This work presented GlycoPy, a CasADi-based Python framework for hierarchical modeling, optimization, and control of bioprocesses. By combining an equation-oriented, object-oriented modeling architecture with the symbolic and differentiable computational capabilities of CasADi, GlycoPy provides a structured and reusable environment for developing large-scale and multiscale process models. In particular, the framework supports hierarchical model composition, numerical and symbolic simulation, parameter estimation, dynamic optimization, and nonlinear model predictive control within a unified workflow.

The case study on monoclonal antibody glycosylation in CHO cell culture demonstrated the main capabilities targeted by the framework. First, the multiscale glycosylation process was naturally organized through hierarchical modeling, which improved the manageability of the large and coupled model. Second, GlycoPy enabled the implementation of a customized differentiable quasi-steady-state simulation strategy, illustrating how symbolic simulation can be used to construct efficient composite simulation workflows for computationally demanding applications. Third, by integrating parameter estimation, simulation, and dynamic optimization, GlycoPy supported adaptive nonlinear model predictive control of the glycosylation process, with the ANMPC results showing clear benefits over open-loop optimization and state NMPC in terms of product-quality improvement while maintaining feasible operation.

Overall, GlycoPy is intended to help bridge the gap between advanced modeling and control algorithms and their practical implementation in complex bioprocess applications. Although the present work focused on bioprocesses, the same framework design is also relevant to other hierarchical process systems. Current limitations include the lack of C-code generation and the present emphasis on shrinking-horizon control and expanding/full-horizon estimation.

Future work will extend the framework to model-based design of experiments, alternative horizon strategies, and more advanced estimation schemes, including simultaneous state and parameter estimation.

**Acknowledgements**

This work was supported by a Project Award Agreement from the National Institute for Innovation in Manufacturing Biopharmaceuticals (NIIMBL), with financial assistance from the U.S. Department of Commerce, National Institute of Standards and Technology (NIST), awards 70NANB17H002 and 70NANB20H037.

**Data Availability**

GlycoPy is released as an open-source package at https://github.com/eaglema/glycopy, and its application in the glycosylation process is released at https://github.com/eaglema/glycosylation.